\documentclass[]{emulateapj}
\usepackage{graphicx}

\usepackage{hhline}
\usepackage{xcolor}
\usepackage{amssymb, amsmath}

\newcommand{\ns}{n_{\rm sat}}
\newcommand{\Ms}{M_{\odot}}

\newcommand{\leff}{\widetilde\Lambda}
\newcommand{\Mc}{\mathcal{M}_c}
\newcommand{\hc}{\hbar c}

\begin{document}

\title{Measurement of the nuclear symmetry energy parameters from gravitational wave events}
\author{Carolyn A. Raithel \& Feryal \"Ozel}
\affiliation{Department of Astronomy and Steward Observatory, University of Arizona, 933 N. Cherry Avenue, Tucson, Arizona 85721, USA}

\begin{abstract}
The nuclear symmetry energy plays a role in determining both the nuclear properties of terrestrial matter as well as the astrophysical properties of neutron stars. The first measurement of the neutron star tidal deformability, from gravitational wave event GW170817, provides a new way of probing the symmetry energy. In this work, we report on new constraints on the symmetry energy from GW170817. We focus in particular on the low-order coefficients: namely, the value of the symmetry energy at the nuclear saturation density, $S_0$, and the slope of the symmetry energy, $L_0$. We find that the gravitational wave data are relatively insensitive to $S_0$, but that they depend strongly on $L_0$ and point to lower values of $L_0$ than have previously been reported, with a peak likelihood near $L_0\sim20$~MeV. Finally, we use the inferred posteriors on $L_0$ to derive new analytic constraints on higher-order nuclear terms.
\end{abstract}

\maketitle

\section{Introduction}

Determining the nuclear symmetry energy is one of the main goals of modern nuclear physics. The symmetry energy, which characterizes the difference in energy between pure neutron matter and matter with equal numbers of protons and neutrons, is typically represented as a series expansion in density, with coefficients that represent the value of the symmetry energy at the nuclear saturation density, $S_0$, the slope, $L_0$, the curvature, $K_{\rm sym}$, and the skewness $Q_{\rm sym}$, as well as higher-order terms. The symmetry energy is one of the two main components in nuclear formulations of the dense-matter equation of state (EOS); the other being the energy of symmetric matter,  which can similarly be broken down into nuclear expansion terms. 

Of these expansion terms, only the low-order parameters can be experimentally constrained, as a result of the limited densities and energies that can be reached in laboratory-based experiments. For example, experimental constraints on $S_0$ and $L_0$ have been inferred by fitting nuclear masses, by measuring the neutron skin thickness, the giant dipole resonance, and electric dipole polarizability of $^{208}$Pb, and by observing isospin diffusion or multifragmentation in heavy ion collisions \citep{Tsang2012,Lattimer2013,Oertel2017}. However, there exist only limited experimental constraints on $K_{\rm sym}$ and no direct constraints on $Q_{\rm sym}$ \citep{Lattimer2013}.

The symmetry energy also plays a key role in a number of astrophysical phenomena, from determining the neutron star radius \citep{Lattimer2001}, to affecting the gravitational wave emission during neutron star mergers (e.g., \citealt{Fattoyev2013}), r-process nucleosynthesis in merger ejecta \citep{Nikolov2011}, and the outcomes of core-collapse supernovae (e.g., \citealt{Fischer2014}). 
In the new gravitational wave era, measurements of the tidal deformability of neutron stars offer a promising way to observationally constrain the symmetry energy. The first detection of gravitational waves from a neutron star-neutron star merger, GW170817, constrained the effective tidal deformability of the binary system to $\leff \lesssim 900$  \citep{Abbott2017a}. Subsequent work refined these constraints to $\leff = 300 \substack{^{+420}_{-230}}$ (90\% highest posterior density) for a system with chirp mass $\Mc=1.186\substack{^{+0.001}_{-0.001}}~\Ms$, for low-spin priors \citep{Abbott2019}.\footnote{Throughout this paper, we will exclusively use the low-spin prior results for GW170817, as is most relevant for binary neutron stars in our Galaxy.} 

Already, several analyses have set initial constraints on nuclear parameters using the tidal deformability of GW170817. \cite{Malik2018} found evidence of correlations between linear combinations of nuclear parameters and the neutron star radius, tidal love number, and tidal deformability, for a wide range of EOS. They used the inferred bounds on $\Lambda_{1.4}$ from GW170817, i.e., the tidal deformability of a 1.4~$\Ms$ neutron star, to constrain the symmetric nuclear parameters as well as $K_{\rm sym}$, for given choices of $L_0$. \cite{Carson2019} expanded this work to include a broader set of EOS and used updated posteriors on $\leff$ to calculate the posteriors for various nuclear parameters. In quoting final constraints on the high-order nuclear parameters, the authors of both studies either limit $L_0$ to a predetermined range or marginalize over priors on $L_0$. However, one might expect $\leff$ to be particularly sensitive to $L_0$, as a result of the direct mapping between $\leff$ and the neutron star radius \citep{Raithel2018,De2018,Raithel2019a} and the tight correlation between the radius and $L_0$ \citep{Lattimer2001}. 

In a more general analysis that allowed for variable $L_0$, \cite{Zhang2019} showed that a precision measurement of $\Lambda_{1.4}$ maps to a plane of constraints on $L_0$, $K_{\rm sym}$, and $Q_{\rm sym}$. They found that the tidal deformability is sensitive to the higher-order symmetry terms ($K_{\rm sym}$ and $Q_{\rm sym}$), and conclude that there is no unique mapping between $\Lambda_{1.4}$ and $L_0$. \cite{Krastev2018} extended this work and showed that the mapping gets even more complicated when the isotriplet/isosinglet interaction is allowed to vary. They found that assuming different density-dependences of the symmetry energy can result in identical values of the $\Lambda_{1.4}$, implying that there can be no one-to-one mapping between the tidal deformability and individual nuclear parameters. 

While there may be no unique mapping of $\Lambda$ to individual nuclear parameters when the parameters are allowed to vary fully independently, we find that a more restricted parameter space is often sufficient to reproduce a wide range of EOS. With a well-motivated parameter reduction, we will show that it becomes possible to directly map from the tidal deformability to nuclear parameters.

In this paper, we introduce a framework to reduce the allowed space of nuclear parameters and we show that constraints can, indeed, be placed directly on the slope of the symmetry energy. Our method does not rely on priors from nuclear experiments. We only assume that the density-dependence of the EOS can be represented with a single-polytrope around the nuclear saturation density, which decreases the parameter space significantly. This dimension-reduction allows us to map from observed constraints on $\leff$ directly to $L_0$, independently of any nuclear priors. We find that the tidal deformability is relatively insensitive to $S_0$, as has been assumed in the above analyses. However, we find that $\leff$ is quite sensitive to $L_0$ and that GW170817 implies a relatively small value of $ 9.0 \lesssim L_0 \lesssim 65.4$~MeV, with a most likely value of $L_0\approx22.5$~MeV. These constraints are approximate, but they point to values of $L_0$ that are significantly lower than those inferred from nuclear physics experiments and theory. Finally, we  use the inferred posterior on $L_0$ to analytically constrain combinations of the higher-order nuclear terms. We find that combinations of $K_{\rm sym}$ and $K_0$ can be constrained by GW170817, and that combinations of $K_{\rm sym}$, $Q_0$, and $Q_{\rm sym}$ can also be constrained.

We start in $\S$\ref{sec:nuclexp} with an overview of the nuclear EOS formalism that we will use in this paper. We introduce our polytropic approximation of these EOS in $\S$\ref{sec:poly}. In $\S$\ref{sec:leff}, we map the measurement of $\leff$ from GW170817 to posteriors over $L_0$. Finally, in $\S$\ref{sec:highorder}, we use the posterior on $L_0$ to constrain linear combinations of the higher-order nuclear parameters.

\section{Nuclear expansion of the equation of state}
\label{sec:nuclexp}

We start by introducing the EOS formalism that we will use to connect the tidal deformability from a gravitational wave event to nuclear parameters. As discussed in the introduction, we use this standard formalism to decompose the EOS into a symmetric matter part and the symmetry energy, which we can generically write as
\begin{equation}
E_b(n, Y_p) = E_0(n) + E_{\rm sym}(n)(1-2Y_p)^2,
\label{eq:Eexp}
\end{equation}
where $E_b(n,Y_p)$ is the energy per baryon for a given density $n$ and proton fraction $Y_p$, $E_0(n)$ is the energy of symmetric matter, and $E_{\rm sym}(n)$ is the symmetry energy.

We represent the symmetric energy term with a series expansion and keep terms to third order, i.e.,  
\begin{equation}
E_0(n) = B_0 + \frac{K_0}{18} u^2 + \frac{Q_0}{162} u^3 + \mathcal{O}(u^4),
\end{equation}
where the expansion is performed around the nuclear saturation density, $\ns$, and  $u \equiv (n/\ns)-1$. Here, $B_0$ is the bulk binding energy of symmetric matter at $\ns$, and $K_0$ and $Q_0$ represent the incompressibility and skewness of symmetric matter. 

Similarly, we expand the symmetry energy around $u$ and write
\begin{equation}
E_{\rm sym}(n) =  S_0 + \frac{L_0}{3} u + \frac{K_{\rm sym}}{18} u^2 + \frac{Q_{\rm sym}}{162}u^3 + \mathcal{O}(u^4),
\end{equation}
where $S_0$ represents the symmetry energy at $\ns$ and $L_0$, $K_{\rm sym}$ and $Q_{\rm sym}$ give the slope, curvature, and skewness of the symmetry energy, respectively.

Such expansions are commonly used in representing neutron star matter because the coefficients can be linked to nuclear physics parameters near the saturation density. While experimental constraints on certain of these parameters exist, in order to be as general as possible, we will only assume knowledge of the bulk binding energy term and fix it to $B = -15.8$ MeV \citep{Margueron2018a}. We will leave the remaining six parameters ($K_0, Q_0, S_0, L_0, K_{\rm sym}, Q_{\rm sym}$) free.

We can convert from the energy per particle to the pressure using the standard thermodynamic relation,
\begin{equation}
P(n,Y_p) = n^2 \left\{ \frac{\partial[ E_b(n,Y_p) + E_e(n,Y_p)] }{\partial n} \right\} \biggr\rvert_{Y_p,S},
\end{equation}
where $S$ is the entropy, and we have formally included the electron contribution to the total energy, $E_e(n,Y_p)$. However, for the current analysis, we neglect the contribution of electrons and assume that the total energy is dominated by the baryons.

The pressure for our nuclear expansion is then
\begin{multline}
\label{eq:Pexp1}
P(n,Y_p) =  \left( \frac{ n^2}{3 n_{\rm sat}} \right)  \times \\
 \left[ \frac{K_0}{3} u + \frac{Q_0}{18} u^2 + \left( L_0 + \frac{K_{\rm sym}}{3} u + \frac{Q_{\rm sym}}{18} u^2 \right)(1-2Y_p)^2 \right].
\end{multline}

For a cold star in $\beta$-equilibrium, the proton fraction is uniquely determined by the density and the symmetry energy, according to
\begin{equation}
\label{eq:Ypfull}
\frac{Y_p}{(1-2Y_p)^3} = \frac{64 E_{\rm sym}(n)^3}{3\pi^2 n (\hbar c)^3}
\end{equation}
where $\hbar$ is the Planck constant and $c$ is the speed of light. (For a derivation of this relationship and an analytic solution for $Y_p$, see Appendix A of \citealt{Raithel2019}). 

In order to simplify the subsequent calculations, we perform an additional series expansion on the neutron excess parameter and define
\begin{equation}
\label{eq:Ypexp}
(1-2Y_p)^2 \approx a + b u + c u^2 + \mathcal{O}(u^3),
\end{equation}
keeping terms up to second-order, as in eq.~(\ref{eq:Pexp1}). The coefficients of this expansion depend on the symmetry energy parameters of up to the same order, i.e., $a = a(S_0),  b = b(S_0, L_0)$, and $c = c(S_0, L_0, K_{\rm sym})$.\footnote{We provide a \texttt{Mathematica} notebook to calculate these coefficients, along with corresponding \texttt{C} routines, at \texttt{https://github.com/craithel/Symmetry-Energy}.} Thus, keeping terms to second order, we can write the nuclear expansion of the pressure as
\begin{multline}
\label{eq:Pexp}
P(n,Y_p) = \left( \frac{ n^2}{3 n_{\rm sat}} \right) 
 \left\{ a L_0 + \left( b L_0 + \frac{K_0 + a K_{\rm sym}}{3} \right) u  \right. \\
 \left. + \left(c L_0 + \frac{b K_{\rm sym}}{3} + \frac{Q_0 + a Q_{\rm sym}}{18}\right) u^2 \right\}.
\end{multline}

\section{Polytropic approximation}
\label{sec:poly}
While the expansion derived in $\S$\ref{sec:nuclexp} is useful for its direct connection to nuclear parameters, it is also complicated. The pressure of eq.~(\ref{eq:Pexp}) depends on 6 nuclear parameters:  $S_0$, $K_0$, $Q_0$, $L_0$, $K_{\rm sym}$, and $Q_{\rm sym}$. However, many studies have shown that a wide range of EOS can be approximated with piecewise polytropic parametrizations (e.g., \citealt{Read2009,Ozel2009,Steiner2010,Raithel2016}). The pressure of a single-polytrope is given by
\begin{equation}
P(n) = K_{\rm poly} n^{\Gamma},
\label{eq:Ppoly}
\end{equation}
where the polytropic constant $K_{\rm poly}$ and index $\Gamma$ are free parameters. The possibility of modeling the EOS with a few number of polytropes motivated us to explore whether the pressure in eq.~(\ref{eq:Pexp}) truly depends on all six nuclear parameters independently, or whether, as we will show, the parameter space can be further restricted.  In this section, we will show that modeling the full pressure of eq.~(\ref{eq:Pexp}) with a single polytrope near the nuclear saturation density reasonably captures the density dependence. We will then use this simplified model to derive constraints on nuclear parameters using data from GW170817.

Our goal is to approximate the nuclear expansion pressure of eq.~(\ref{eq:Pexp}) with the polytropic pressure of eq.~(\ref{eq:Ppoly}). We require that these two expressions match at $n_{\rm sat}$ and then extrapolate to higher densities using the polytropic index. This requirement uniquely determines the polytropic constant, so that our simplified nuclear pressure can be written as
\begin{equation}
\label{eq:Pmodel}
P(n) = \frac{a L_0 n_{\rm sat} }{3} \left( \frac{n}{n_{\rm sat}} \right)^{\Gamma}.
\end{equation}
At low densities of $n \le 0.5~n_{\rm sat}$, we fix the EOS to the nuclear EOS SLy \citep{Douchin2001}. For $0.5~n_{\rm sat} \le n < n_{\rm sat}$, we perform a power-law interpolation, to ensure matching between SLy and the polytropic approximation.

In order to test whether this simplified model of the pressure reasonably captures the density-dependence of the full nuclear expansion, we generate a sample of 1,000 test EOS using eq.~(\ref{eq:Pexp}). The EOS are created by drawing independent values of each of the six nuclear parameters ($K_0, Q_0, S_0, L_0, K_{\rm sym},$ and $Q_{\rm sym}$) from the experimentally-constrained distributions reported in Table I of \cite{Margueron2018} (a similar approach was taken in \citealt{Carson2019}). We exclude any EOS that become hydrostatically unstable or that have superluminal sound speeds across a density range of $n\sim0.01-10~n_{\rm sat}$. Additionally, we require that the analytic expression for $Y_p$ derived from eq.~(\ref{eq:Ypfull}) be positive and less than 0.5 (i.e., neutron rich) across the same density range.

\begin{figure}[ht]
\centering
\includegraphics[width=0.48\textwidth]{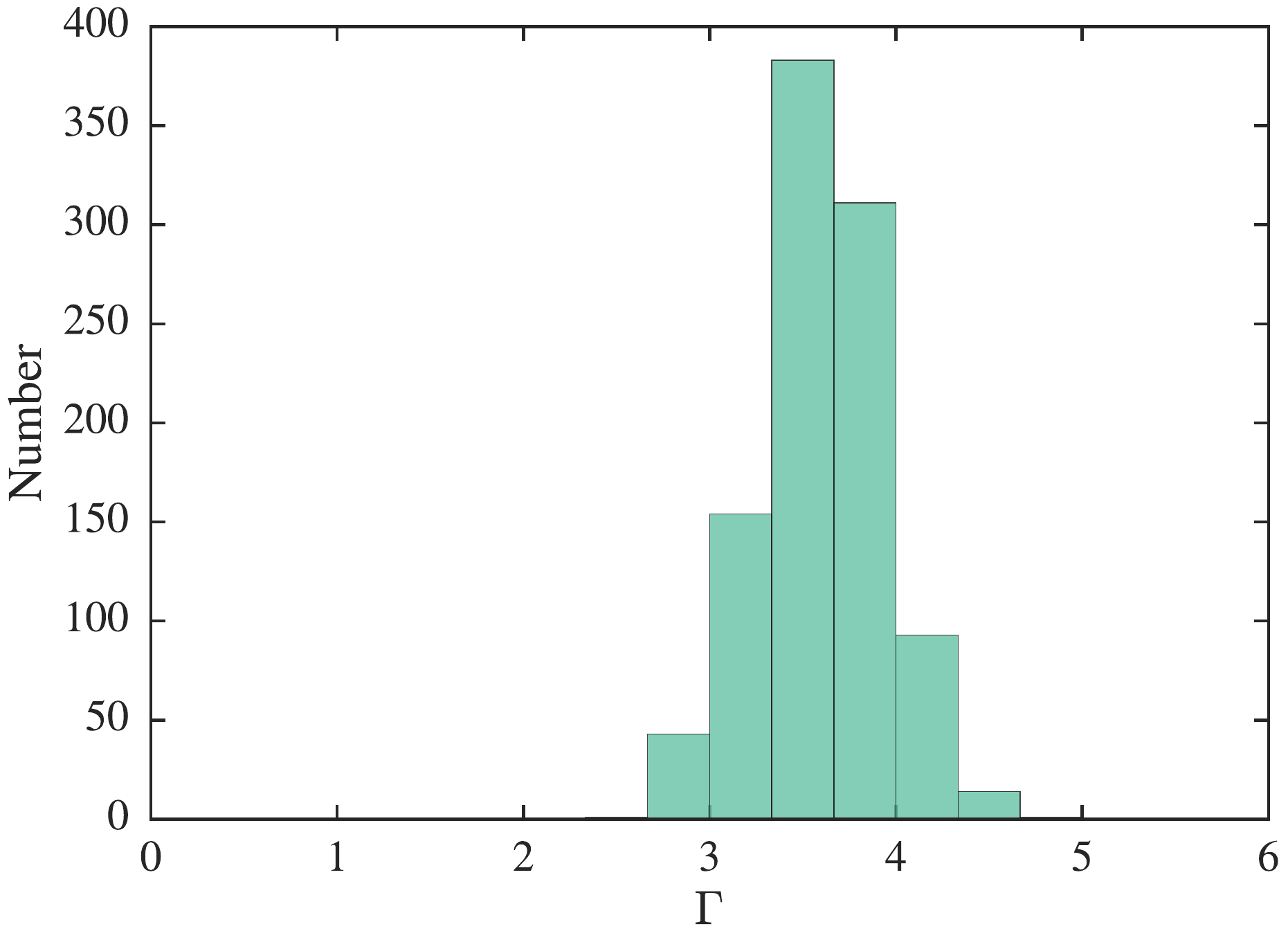}
\caption{\label{fig:gamma} Distribution of polytropic indices fit to a sample of 1,000 nuclear expansion EOS using eq.~(\ref{eq:Pmodel}). Each EOS was constructed using eq.~(\ref{eq:Pexp}) with the six nuclear parameters, ($K_0, Q_0, S_0, L_0, K_{\rm sym},$ and $Q_{\rm sym}$),  independently drawn from experimentally-constrained distributions. We find that nearly all of the EOS can be with with $\Gamma=3-4$, while the most common $\Gamma$ is $\sim3.5$} 
\end{figure}

\begin{figure}[ht]
\centering
\includegraphics[width=0.48\textwidth]{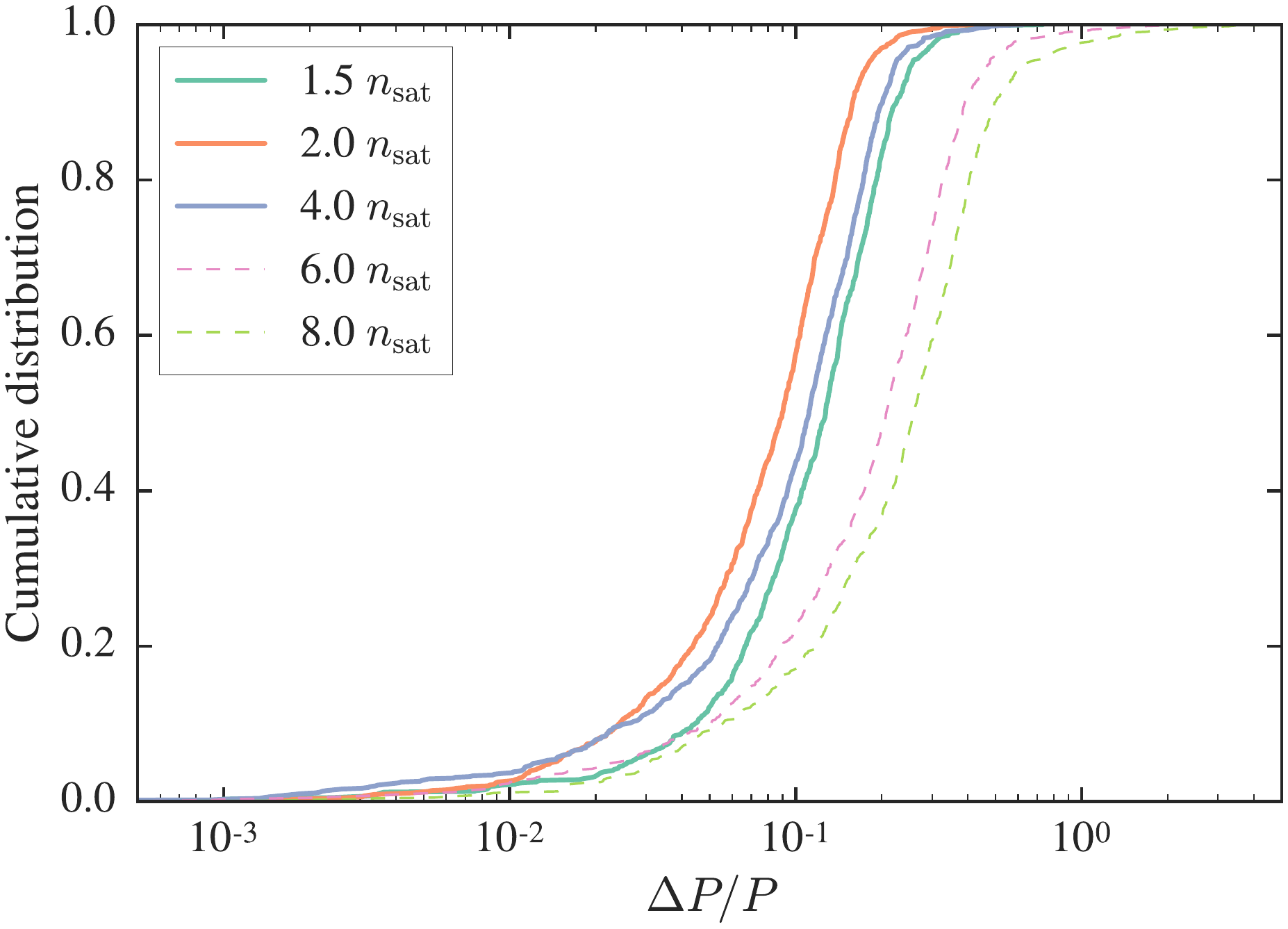}
\caption{\label{fig:resids} Cumulative distribution of residuals between the pressure of the full nuclear expansion in eq.~(\ref{eq:Pexp}) and our single-polytrope approximation of eq.~(\ref{eq:Pmodel}), calculated at various fiducial densities (shown in the different colors). We find that the single-polytrope approximation reasonably captures the overall density-dependence of the pressure. At densities of $2~\ns$, which are expected to determine the neutron star radius and hence the tidal deformability, the errors of our polytropic approximation are $\lesssim$15\% for 90\% of the EOS in our sample.} 
\end{figure}

\begin{figure*}[ht]
\centering
\includegraphics[width=\textwidth]{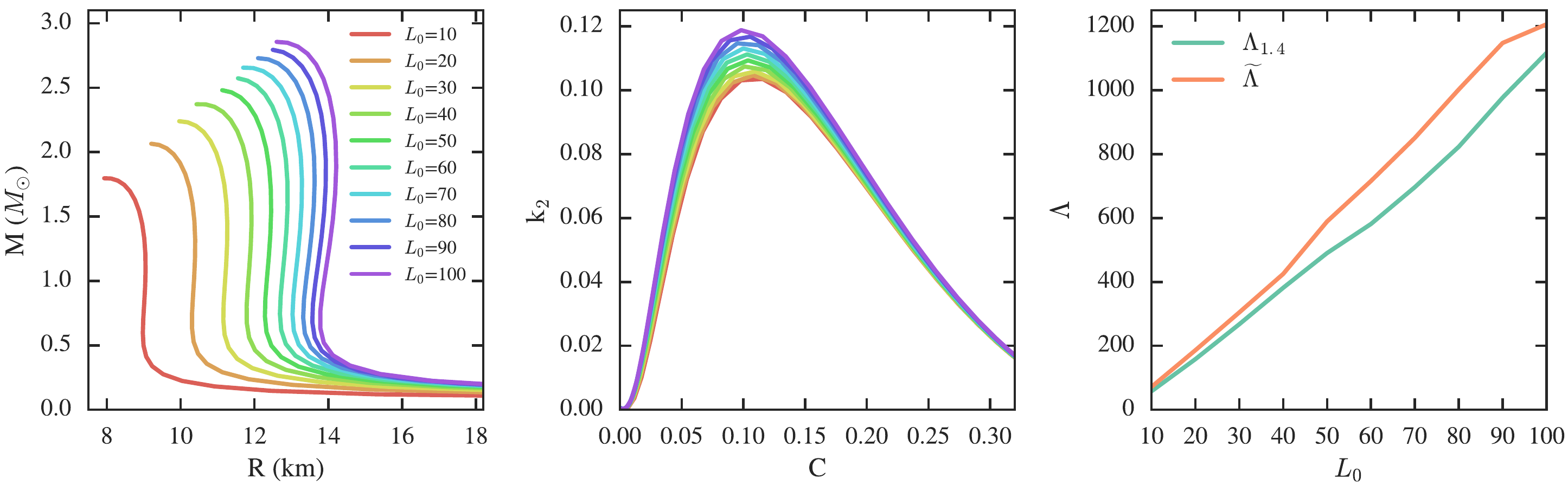}
\caption{\label{fig:k2} Left:  Mass-radius curves for our polytropic approximation with varying values for $L_0$ (in MeV). Middle:  Tidal apsidal constants for the same EOS, as a function of stellar compactness ($C = Gm/R c^2$). Right:  Tidal deformability as a function of $L_0$. In all three panels, we have fixed $S_0$ to 32~MeV and $\Gamma=3.5$. In the right panel, $\leff$ is calculated assuming the $q=0.87$ and $\Mc=1.186~\Ms$, as was observed for GW170817. Within the polytropic approximation, we find that smaller values of $L_0$ correspond both to smaller radii and to smaller values of  $k_2$, resulting in a smaller tidal deformability. } 
\end{figure*}

We fit each EOS in our sample with the simplified pressure model of eq.~(\ref{eq:Pmodel}), fixing $L_0$ and $S_0$ to their drawn values. We perform the fit across the density range $n=1-3\ns$, in order to most strongly weight the density regime which is responsible for determining the neutron star radius, and hence the effective tidal deformability \citep{Lattimer2001,Raithel2018}. We show the resulting distribution of $\Gamma$ values in Fig.~\ref{fig:gamma}. We find that nearly all of the EOS constructed with the nuclear expansion formalism can be represented with a polytropic index of $\Gamma=3-4$, with a most common fit value of $\Gamma\sim3.5$. Moreover, we find that the residuals between the full nuclear expansion EOS and our polytropic approximation are small. Figure~\ref{fig:resids} shows the cumulative distribution of the residuals from the EOS fits at a range of densities. The residuals are smallest at low densities, where the tidal deformability is expected to be determined and where the nuclear expansion formalism still applies. At $2~n_{\rm sat}$, the error introduced by our polytropic approximation is  $\lesssim$15\% for 90\% of the EOS sample. For completeness, we also show in Fig.~\ref{fig:resids} the residuals at core densities of $6-8~n_{\rm sat}$  (thin, dashed lines), even though the symmetry energy expansion is expected to break down at the these high densities. At $8~n_{\rm sat}$, the residuals of our approximation are still $\lesssim$~50\% for 90\% of the sample.

We, therefore, find that the single-polytrope approximation reasonably recreates $P(n)$ for most combinations of the nuclear parameters. While this approximation is not exact, it is a useful technique that will allow us to explore the parameter-dependence of $\leff$ in a new way. Our simplified model depends only on $S_0$ and $L_0$, thereby reducing a six-dimensional parameter space to two dimensions. This will allow us to directly map from $\leff$ to $S_0$ and $L_0$, without requiring us to fix or marginalize over the higher-order terms.

\section{Relating the tidal deformability to the symmetry energy}
\label{sec:leff}

\begin{figure*}[ht]
\centering
\includegraphics[width=\textwidth]{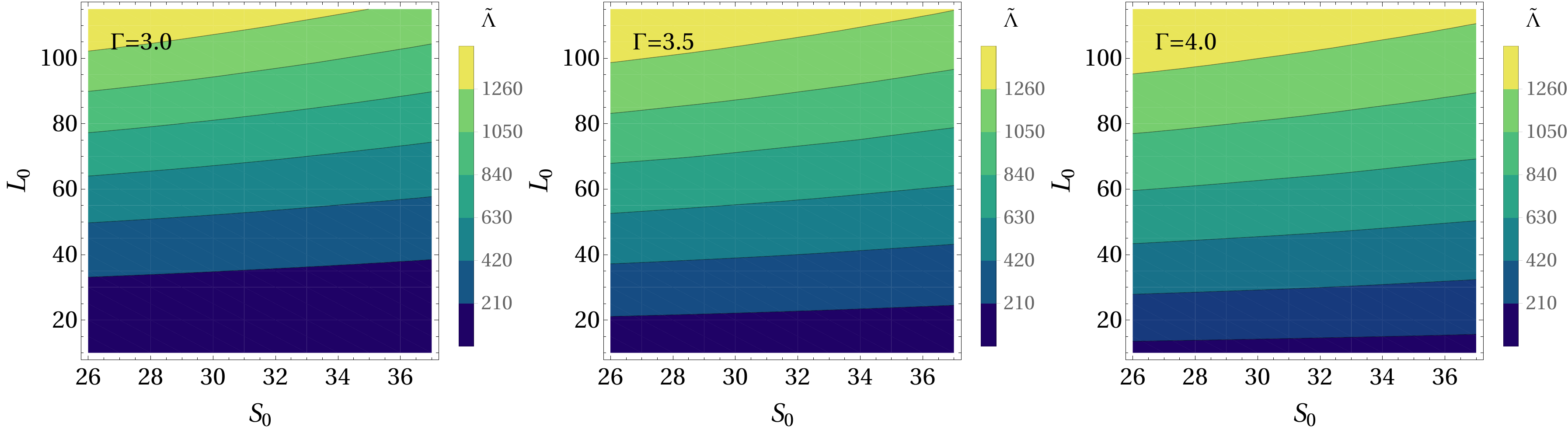}
\caption{\label{fig:S0L} The effective tidal deformability of the binary system, as a function of  $S_0$ and $L_0$. We calculate $\leff$ using the polytropic approximation of the nuclear EOS, shown in eq.~(\ref{eq:Pmodel}). From left to right, the polytropic index is fixed to $\Gamma=3,$ 3.5, or 4. In all panels, we fix the chirp mass and mass ratio to their central values of $q=0.87$ and $\Mc=1.186~\Ms$ for GW170817. We find that  $\leff$ is only weakly dependent on $S_0$, but that it is quite sensitive to $L_0$. The constraints on $\leff = 300~(+420/-230)$ from GW170817 \citep{Abbott2019} point to relatively small values of $L_0$.} 
\end{figure*}

Using the framework for pressure introduced in $\S$\ref{sec:poly}, we can now connect the observed constraints on $\widetilde{\Lambda}$ from a gravitational wave event to nuclear parameters. We start with the general expression for the tidal deformability of a single star,
\begin{equation}
\label{eq:lambda}
\Lambda_i = \frac{2}{3} k_2 \left( \frac{Gm_i}{R_ic^2} \right)^{-5},
\end{equation}
where $m_i$ is the mass of the star, $R_i$ is the stellar radius, and, following the convention of \citet{Flanagan2008}, we call $k_2$ the tidal apsidal constant. The tidal apsidal constant depends both on the compactness of the star, as well as the overall density gradient of the particular EOS \citep{Hinderer2008, Hinderer2010, Postnikov2010}. 

We follow the method outlined in \citet{Hinderer2010} for constructing a set of augmented Oppenheimer-Volkoff equations. We integrate these stellar structure equations to calculate the stellar mass, radius, and tidal apsidal constant for a given central density. We then compute the effective tidal deformability of the binary system as
\begin{equation}
\leff = \frac{16}{13} \frac{ (m_1 +  12 m_2) m_1^{4} \Lambda_{1} + (m_2 + 12 m_1) m_2^{4} \Lambda_{2}}{(m_1+m_2)^5},
\label{eq:leff}
\end{equation}
where the subscripts indicate the component stars in the binary system.

We show the effect of $L_0$ on each of these stellar properties in Fig.~\ref{fig:k2}. For demonstrative purposes, in this figure we have fixed $S_0$=32~MeV and $\Gamma=3.5$, as well as the component masses for calculating $\leff$ (see below for the effect of varying each of these assumptions). We show the mass-radius relations for a variety of $L_0$ in the left panel of Fig.~\ref{fig:k2}.  The middle panel panel shows the tidal apsidal constant as a function of the stellar compactness, for the same set of $L_0$ values. We find that smaller values $L_0$ lead to both smaller radii and to smaller tidal apsidal constants. Both of these trends act to reduce the tidal deformability of the star (see eq.~\ref{eq:lambda}), as shown in the right panel of Fig.~\ref{fig:k2}. We find that the dependence on $L_0$ persists both for $\Lambda_{1.4}$ and for the binary tidal deformability, $\widetilde{\Lambda}$. Thus, we expect that the measurement of $\leff$ from a gravitational wave event should have significant constraining power on $L_0$.
 
From eq.~(\ref{eq:leff}) and Fig.~\ref{fig:k2}, it is clear that $\leff$ depends on the stellar masses and radii, with an additional dependence on the EOS through $k_2$.
By introducing the chirp mass, 
\begin{equation}
\label{eq:mc}
\Mc = \frac{(m_1 m_2)^{3/5} }{(m_1 + m_2)^{1/5}} = m_1 \frac{q^{3/5}}{(1+q)^{1/5}},
\end{equation}
and the mass ratio $q \equiv m_2/m_1$, we can explicitly write the dependences of $\leff$ as $\leff(\Mc, q, R_1, R_2, \rm{EOS})$. This is a particularly convenient choice because, for gravitational wave events, we expect the chirp mass to be precisely measured and the mass ratio to also be constrained. This was indeed the case for GW170817, for which the the chirp mass was determined to be $\Mc=1.186\substack{+0.001\\-0.001}~\Ms$ and the mass ratio was constrained to $q \in (0.73, 1.00)$ at the 90\% confidence level  \citep{Abbott2019}.

Additionally, we note that, given the masses of each star, the EOS can be used to uniquely determine the corresponding radii.\footnote{While there do exist some EOS for which the mass-to-radius mapping is not unique (notably, the so-called ``twin-stars," which can have identical masses and different radii; see, e.g., \citealt{Glendenning2000}), these EOS have complex structure that cannot be represented with single polytropes. We, therefore, neglect these special cases for the present study.} Within the polytropic approximation of eq.~(\ref{eq:Pmodel}), the EOS depends only on $S_0$, $L_0$, and $\Gamma$, where $\Gamma$ is narrowly constrained to be $\sim3-4$ for a wide range of realistic EOS.  We can, therefore, summarize the dependences of the tidal deformability as $\leff = \leff(\Mc, q, S_0, L_0, \Gamma)$.

We have already shown that $\leff$ depends sensitively on $L_0$ for fixed $S_0$ and $\Gamma$. In order to explore the full, more general dependences of $\leff$, we perform a grid search across the range $S_0 \in (26,38)$~MeV and $L_0 \in (10,120)$~MeV. For each set of values, we construct an EOS according to eq.~(\ref{eq:Pmodel}), fixing the mass ratio to $q=0.7, 0.87,$ or 1.0 and fixing $\Gamma$ to 3, 3.5, or 4. In all cases, we fix the chirp mass to the central value from GW170817 of $1.186~\Ms$. For each combination of parameters, we compute the mass, radius, and tidal apsidal constant by numerically integrating the augmented TOV equations and then compute $\leff$ using eqs.~(\ref{eq:lambda}-\ref{eq:mc}). We show the resulting contours of $\leff$ as a function of $S_0$ and $L_0$ in Fig.~\ref{fig:S0L}. The three panels correspond to three different choices of $\Gamma$, with fixed $q=0.87$. We find that the particular choice of $q$ does not significantly affect these or our later results, so we fix $q$ to the central value of 0.87 from GW170817 for the remainder of this analysis.

We find that $\leff$ is only weakly dependent on $S_0$, especially for smaller values of $\leff$, as are preferred by the current gravitational wave data. In contrast, $\leff$ depends quite sensitively on $L_0$. We, therefore, focus on $L_0$ in the following analysis and fix $S_0$ to a characteristic value of 32~MeV \citep{Li2013,Oertel2017}.

This final simplification renders $\leff$ as a function only of $L_0$, for fixed $\Gamma$. We can, therefore, transform the measured posterior on $\leff$ to a posterior on $L_0$, according to
\begin{equation}
\mathcal{P}(L_0) = \mathcal{P}(\leff) \left( \frac{\partial \leff}{\partial L_0} \right),
\end{equation}
where we calculate the Jacobian term numerically. 

\begin{figure}[ht]
\centering
\includegraphics[width=0.45\textwidth]{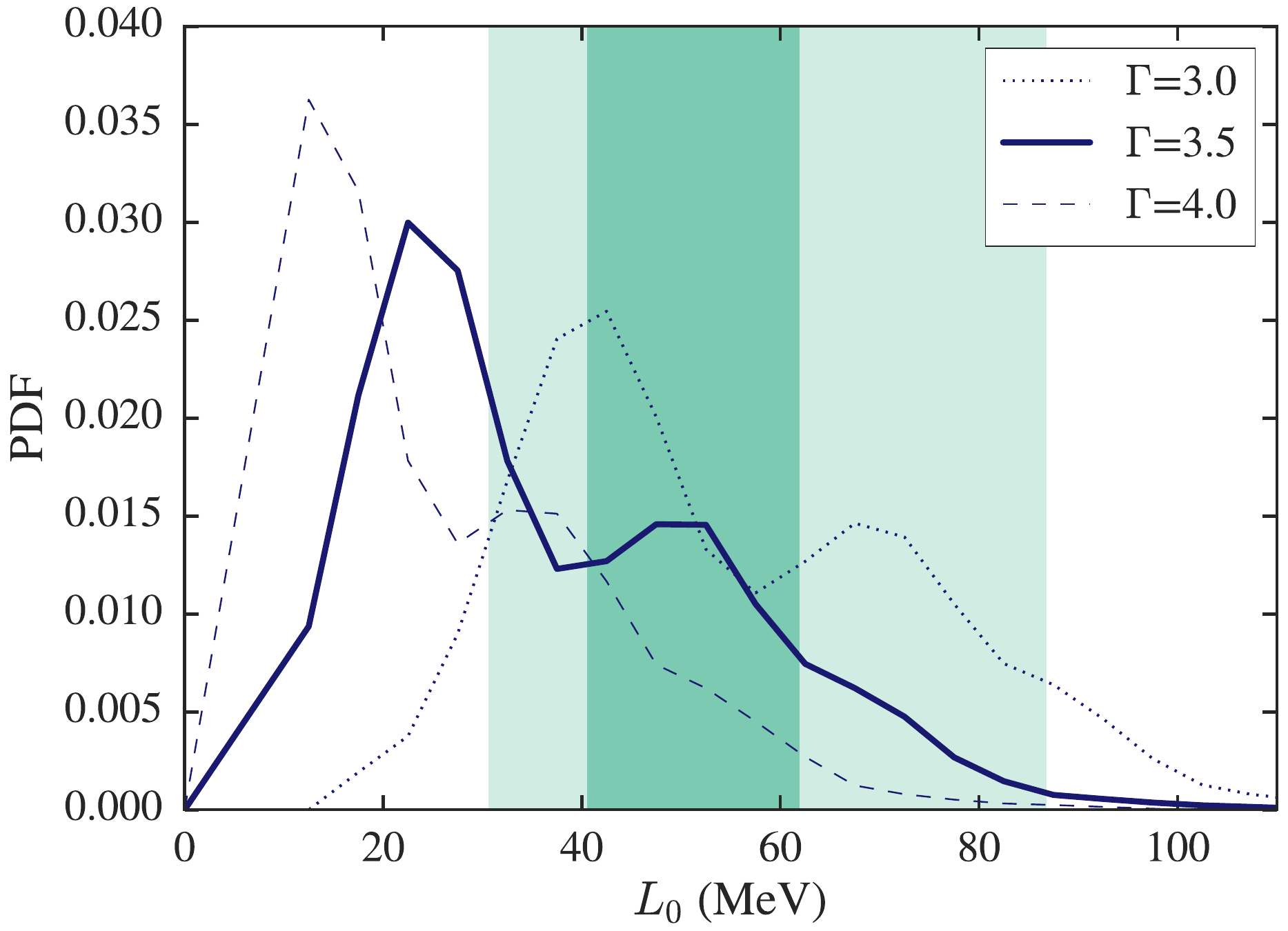}
\caption{\label{fig:margL} One-dimensional posterior in $L_0$, from GW170817 for $q=0.87$ and $S_0$=32~MeV, for three choices of $\Gamma$. The dark and light green bands show the combined constraints on $L_0$ from previous neutron star observations, nuclear experiments, and theory, as calculated in \cite{Lattimer2013} and \cite{Oertel2017}, respectively. We find that the gravitational wave data point towards smaller values of $L_0$ than these previous studies have found.} 
\end{figure}

We show the resulting one-dimensional posteriors on $L_0$ in Fig.~\ref{fig:margL}. Figure~\ref{fig:margL} also shows two current sets of constraints on $L_0$, in dark and light green from \cite{Lattimer2013} and \cite{Oertel2017}, respectively, which are based on a combination of astrophysical observations of neutron stars, nuclear experiments, and theory. Earlier constraints on $43 < L_0 < 52$~MeV (68\% confidence) were calculated using neutron star radii alone \citep{Steiner2012}. On the other hand, theoretical calculations of the neutron matter EOS using quantum Monte Carlo methods \citep{Gandolfi2012} or chiral effective field theory \citep{Hebeler2013} produce comparable constraints, of $L_0=31.3-63.6$~MeV and $L_0=32.4-57.0$~MeV, respectively. We use the summary results from \cite{Lattimer2013} and \cite{Oertel2017} to encompass these theoretical, observational, and experimental constraints.

We find that the gravitational wave data imply smaller values of $L_0$ than these previous studies have found. In particular,  for $\Gamma=3.5$, we find a 90\% highest-posterior density interval of $9.0 < L_0 <  65.4$~MeV, with a peak likelihood at $L_0\approx22.5$~MeV. There is a small correlation between the choice of $\Gamma$ and the inferred constraints on $L_0$, with choices of larger values of $\Gamma$ leading to lower values of $L_0$.  Nevertheless, for $\Gamma=3.5$ or 4, the peak likelihoods in $L_0$ lie outside the allowed constraints from both \cite{Lattimer2013} and \cite{Oertel2017}. For $\Gamma=3$, the peak likelihood falls at the lower limit of the constraint from \cite{Lattimer2013}.

Several recent studies connecting GW170817 to the nuclear EOS have either restricted $L_0$ to similar priors or marginalized over them \citep{Malik2018,Carson2019}. While the posteriors on $L_0$ presented here are not exact, they do suggest that GW170817 points toward small values of $L_0$ that may be in tension with such priors. We, therefore, conclude that it is important to explore the dependence of gravitational wave data on $L_0$ directly, in order to gain new information on $L_0$ itself as well as to avoid biasing the interpretation of higher-order parameters.

\section{Constraints on higher-order nuclear parameters}
\label{sec:highorder}
In $\S$\ref{sec:leff}, we showed that GW170817 directly maps to constraints on $L_0$ using our polytropic approximation of the nuclear expansion. In this section, we turn to the higher-order nuclear terms. In particular, we will show that by taking the polytropic approximation in eq.~(\ref{eq:Pmodel}), we can place constraints on the allowed combinations of the remaining four nuclear parameters.

We start by equating the polytropic approximation and the full nuclear expansion, and match terms of equivalent order, i.e., 
\begin{multline}
\label{eq:orders}
a L_0 (u+1)^{\Gamma-2} = 
a L_0 +
\left( b L_0 + \frac{K_0 + a K_{\rm sym}}{3}  \right)  u   \\
+ \left( c L_0 + \frac{b K_{\rm sym}}{3} + \frac{Q_0 + a Q_{\rm sym}  }{18} \right) u^2.
\end{multline}

For this expression to be true at all densities, the terms of equivalent order must all sum to zero. For example, setting $\Gamma=3$ implies the constraints 
\begin{subequations}
\begin{equation}
\label{eq:g3_1}
 (a - b) L_0 - \frac{K_0 + a K_{\rm sym}}{3}  = 0   
\end{equation}
\begin{equation}
\label{eq:g3_2}
 c L_0 + \frac{ b K_{\rm sym}}{3} + \frac{Q_0 + a Q_{\rm sym}}{18} = 0.
\end{equation}
\end{subequations}
Likewise, setting $\Gamma=4$ implies
\begin{subequations}
\begin{equation}
\label{eq:g4_1}
(2 a - b) L_0 - \frac{K_0 + a K_{\rm sym}}{3} = 0   
\end{equation}
\begin{equation}
\label{eq:g4_2}
(c-a) L_0 +\frac{ b K_{\rm sym}}{3}+ \frac{Q_0 + a Q_{\rm sym} }{18}  = 0.
\end{equation}
\end{subequations}
For $\Gamma=3.5$, we introduce one final series expansion on the left-hand side of eq.~(\ref{eq:orders}) to simplify $(u+1)^{1.5} \approx 1 + (3/2)u + (3/8)u^2 + \mathcal{O}(u^3)$. Using this approximation, we can again require terms of the same order to sum to zero, and we find
\begin{subequations}
\begin{equation}
\label{eq:g35_1}
\left(\frac{3 a}{2} - b \right) L_0 - \frac{K_0 + a K_{\rm sym}}{3}  = 0   
\end{equation}
\begin{equation}
\label{eq:g35_2}
\left(c-\frac{3a}{8}\right) L_0 +\frac{ b K_{\rm sym}}{3}+ \frac{Q_0 + a Q_{\rm sym} }{18}  = 0,
\end{equation}
\end{subequations}
where we recall that $a=a(S_0), b=b(S_0,L_0),$ and $c=c(S_0,L_0,K_{\rm sym})$. 

Thus, for a given choice of $\Gamma$, we have two independent sets of constraints: the first connects the parameter set $\{S_0$, $L_0$, $K_0$, and $K_{\rm sym}\}$, while the second connects $\{S_0$, $L_0$, $K_{\rm sym}$, $Q_0$, $Q_{\rm sym}\}$. In the following, we will use the posterior on $L_0$ from $\S$\ref{sec:leff} to constrain the remaining combinations of higher-order terms using these relationships.

\begin{figure}[ht]
\centering
\includegraphics[width=0.5\textwidth]{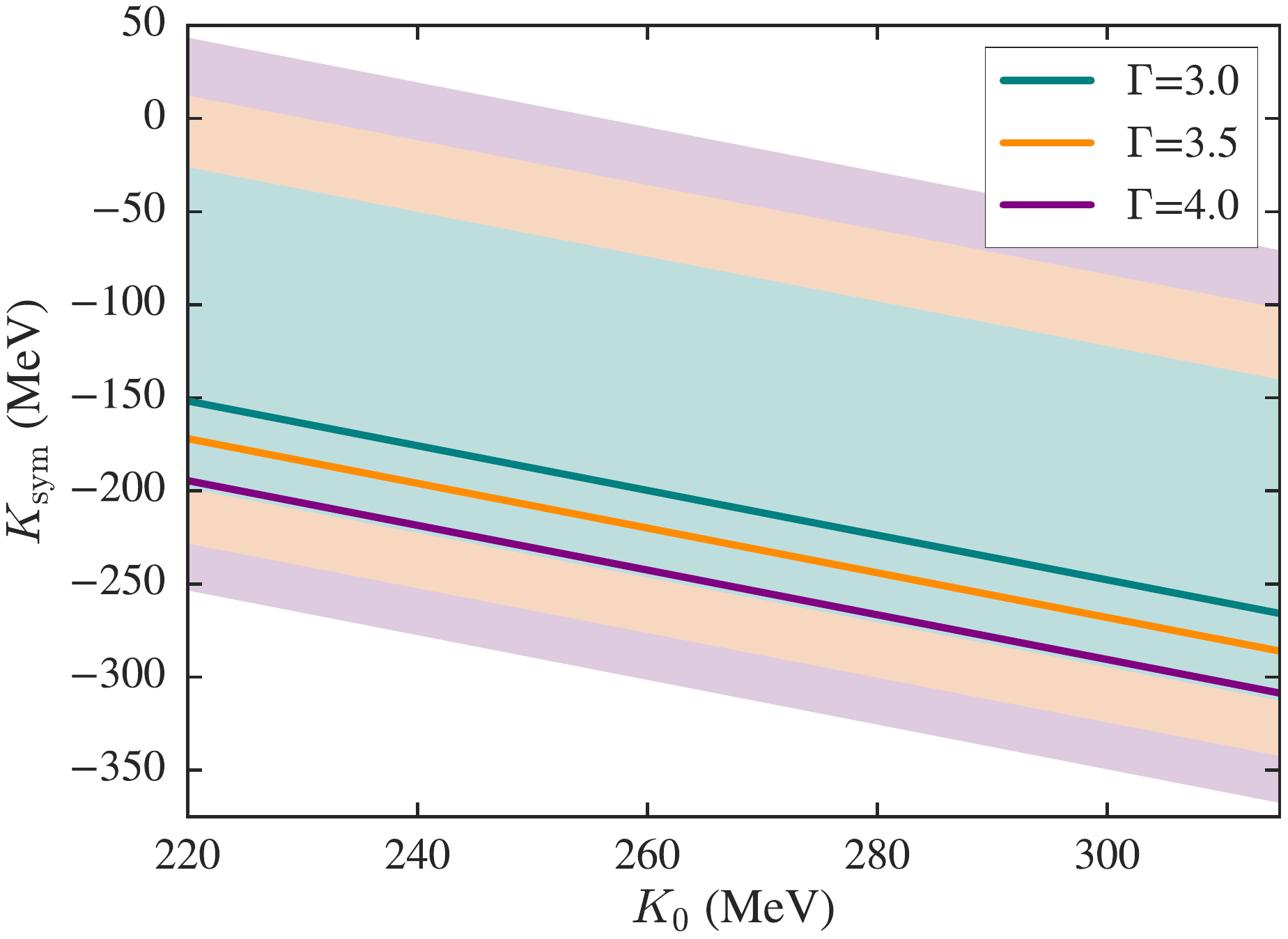}
\caption{\label{fig:KKsym} Two-dimensional constraints on $K_{\rm sym}$ and $K_0$, from GW170817. The shaded regions bound the $K_{\rm sym}-K_0$ space allowed by eqs.~(\ref{eq:g3_1}), (\ref{eq:g4_1}), and (\ref{eq:g35_1}), for $S_0$=32~MeV and $L_0$ corresponding to the 90\% confidence intervals from Fig.~\ref{fig:margL}. The dark, solid lines represent the $K_{\rm sym}-K_0$ relationship corresponding to the most likely value of $L_0$ from Fig.~\ref{fig:margL} for each $\Gamma$. We find that GW17017 constrains $K_{\rm sym}$ to values $\lesssim 70$~MeV, for a wide range $K_0$.} 
\end{figure}

We start with the first set of constraints, on  $\{S_0$, $L_0$, $K_0$, and $K_{\rm sym}\}$. These constraints correspond to eqs.~(\ref{eq:g3_1}), (\ref{eq:g4_1}), and (\ref{eq:g35_1}). As in $\S$~\ref{sec:leff}, we fix $S_0$=32~MeV and find that this choice does not strongly affect the results.  We then use the 90\%-credible interval on $L_0$ from Fig.~\ref{fig:margL} to bound the allowed range of $K_{\rm sym}-K_0$ values. We show the resulting constraints in Fig.~\ref{fig:KKsym} for each $\Gamma$. In this figure, the light shaded regions represent the bounds on the $K_{\rm sym}-K_0$ relationship allowed by the 90\%-credible interval on $L_0$. The dark solid lines indicate the $K_{\rm sym}-K_0$ relationship corresponding to the most likely value of $L_0$, for a given $\Gamma$.

We find that GW170817 places tight constraints on linear combinations of $K_0$ and $K_{\rm sym}$.  For the three values of $\Gamma$, fixing $L_0$ to its maximum likelihood value from Fig.~\ref{fig:margL} yields 
\begin{subequations}
\label{eq:Ksym}
\begin{align}
\Gamma=3.0: \quad &K_{\rm sym} = 112.54 - 1.201 K_0 \\
\Gamma=3.5: \quad &K_{\rm sym} = 92.32 - 1.201 K_0 \\
\Gamma=4.0: \quad &K_{\rm sym} = 69.76 - 1.201 K_0.
\end{align}
\end{subequations}
These equations correspond to the dark, solid lines in Fig.~\ref{fig:KKsym}. The coefficient in front of $K_0$ is the same in all three cases because it is given simply by $1/a$ and hence depends only on $S_0$. The constant term depends on $S_0$, $L_0$, and the coefficients of $L_0$ in eqs.~(\ref{eq:g3_1}), (\ref{eq:g4_1}), and (\ref{eq:g35_1}), and thus varies slightly with the choice of $\Gamma$.

Previous studies have constrained $K_0$ by fitting nuclear models to measurements of the isoscalar giant monopole resonance. Depending on the analysis methods, the results range from quite narrow, $K_0=248\pm8$~MeV \citep{Piekarewicz2004} and $K_0=240\pm20$~MeV \citep{Shlomo2006}, to broader bounds of $K_0 = 250-315$~MeV \citep{Stone2014}. If we take the broadest range of these allowed values and assume $220 < K_0 < 315$ and combine this with our results in Fig.~\ref{fig:KKsym}, we find that $-375 \lesssim K_{\rm sym} \lesssim  45$, at 90\% confidence. This constraint on $K_{\rm sym}$ is broader than, but consistent with previous results, including the constraint of $K_{\rm sym}=-111.8\pm71.3$~MeV that was derived from universal relations between $K_{\rm sym}$ and lower-order expansion terms \citep{Mondal2017}. 

Using the tidal deformability from GW170817, \citet{Malik2018} found constraints of $-112 < K_{\rm sym} <-52$~MeV or $-140 < K_{\rm sym} < 16$~MeV, depending on their choice of prior for $L_0$. In a similar analysis, \citet{Carson2019} derived constraints of $-259 \le K_{\rm sym} \le 32$~MeV, after marginalizing over $L_0$. Our results, which are derived with the polytropic approximation and no priors on $L_0$, are consistent with both of these analyses. However, if we take the $K_{\rm sym}(K_0)$ relationship that corresponds to the maximum likelihood in $L_0$ (i.e., the dark solid lines in Fig.~\ref{fig:KKsym}), we find that the data point to smaller values of $K_{\rm sym}$, below the lower bound from the \citet{Malik2018} study and on the lower end of the \citet{Carson2019} constraints. This is likely a consequence of the fact that these studies both used priors that either forbade or disfavored low values of $L_0$, such as those we find in this paper. 

Finally, we turn to the second set of constraints on the parameters $\{S_0$, $L_0$, $K_{\rm sym}$, $Q_0$, $Q_{\rm sym}\}$. As we did above, we will fix $S_0=32$~MeV and use the maximum likelihood in $L_0$. We can then use  eqs.~(\ref{eq:g3_2}), (\ref{eq:g4_2}), and (\ref{eq:g35_2}) to calculate the relationship between the remaining four parameters. For the most likely value of $L_0$, we find
\begin{subequations}
\begin{align}
\Gamma=3.0:  \quad  &K_{\rm sym} = 107.8 - 1.74 Q_0 - 1.45 Q_{\rm sym}  \\
\Gamma=3.5:  \quad  &K_{\rm sym} = 247.1 - 1.53 Q_0 - 1.27 Q_{\rm sym} 	 \\
\Gamma=4.0: \quad   &K_{\rm sym} = 299.0 - 1.44 Q_0 - 1.20 Q_{\rm sym}.
\end{align}
\end{subequations}

To our knowledge, no nuclear experiments have constrained $Q_0$ or $Q_{\rm sym}$ and only broad theoretical bounds have been calculated. For example, \citet{Zhang2017} found $-800 < Q_0 < 400$~MeV based on analyses of energy density functionals. Nevertheless, future experiments or astrophysical observations may one day provide stricter bounds on $Q_0$ or $Q_{\rm sym}$. Within our polytropic framework, any such measurements can then be used to constrain the correlated parameters using these analytic relationships.

\section{Conclusions}
In this paper, we have introduced a new approximation of the nuclear EOS which allows for a direct mapping from measured $\leff$ constraints to the symmetry energy parameters. We have shown that a wide sample of nuclear EOS can be reasonably represented with a single-polytrope approximation in the density range of interest, which simplifies the EOS to depend only on $S_0$, $L_0$, and $\Gamma$, rather than the full six nuclear parameters. Moreover, we find that $\leff$ is relatively insensitive to $S_0$. With many future gravitational wave detections expected in the coming years, this framework will make it possible to map the gravitational wave event directly to $L_0$ or to combinations of higher-order nuclear parameters. 

With this parameter-space reduction and focusing on the existing measurement of $\leff$ from GW170817, we were able to map from the full posterior on $\leff = 300 \substack{^{+420}_{-230}}$ \citep{Abbott2019} to posteriors on $L_0$. We find that GW170817 points to significantly smaller values of $L_0$ than have been previously been reported, with a peak likelihood of $L_0\sim20$~MeV.
We additionally use these posteriors on $L_0$ to constrain combinations of higher-order nuclear parameters, finding tight constraints on the allowed combinations of $K_{\rm sym}$ and $K_0$, as well as constraints on $K_{\rm sym}, Q_0,$ and $Q_{\rm sym}$.

We note that the final constraints on $L_0$ depend slightly on the choice of $\Gamma$ and, of course, will depend on the robustness of our polytropic approximation. If the true combination of $K_0, Q_0, S_0, L_0, K_{\rm sym},$ and $Q_{\rm sym}$ produce an EOS with significant sub-structure, then our single-polytrope approximation is not the optimal approach. Moreover, if the dense-matter EOS contains a phase transition to quark matter, then the polytropic approximation will be inadequate and, depending on the particular formulation, the relationship between the tidal deformability and $L_0$ may be significantly weaker as well (e.g., \citealt{Zhu2018}).

Nevertheless, the results in this paper indicate that gravitational wave data can significantly constrain the slope of the symmetry energy for nuclear EOS. This is an important point. Previous studies connecting GW170817 to the nuclear EOS have either fixed the allowed range of $L_0$ or marginalized over $L_0$ \citep{Malik2018,Carson2019}, using priors from nuclear physics that we find to be in modest conflict with the values inferred from GW170817. As the LIGO/Virgo team continue to observe new gravitational wave events and further pin down the tidal deformability of neutron stars, it will become increasingly important to develop robust approaches to constrain the nuclear parameters in model-independent ways.
\\\\
\textit{Acknowledgments.} We thank Dimitrios Psaltis, Kent Yagi, Andrew Steiner, and Zack Carson for useful conversations related to this work. This work is supported by NSF Graduate Research Fellowship Program Grant DGE-1746060 and support from NASA grant NNX16AC56G.

\bibliography{gwbib}
\bibliographystyle{apj}
\end{document}